\documentclass[pdflatex,sn-mathphys-num]{sn-jnl}

\usepackage{multirow}%
\usepackage{amsthm}%
\usepackage{mathrsfs}%
\usepackage[title]{appendix}%
\usepackage{xcolor}%
\usepackage{textcomp}%
\usepackage{manyfoot}%
\usepackage{booktabs}%
\usepackage{algorithm}%
\usepackage{algorithmicx}%
\usepackage{algpseudocode}%
\usepackage{listings}%
\usepackage[utf8]{inputenc}
\usepackage{amssymb,amsmath,amsfonts}
\usepackage{graphicx,psfrag,color}
\usepackage{dcolumn}
\usepackage{bbm}
\usepackage{mathbbol}
\usepackage{braket}
\usepackage{times}
\usepackage{graphicx}
\usepackage{verbatim}
\usepackage{hyperref}
\usepackage{ragged2e}
\usepackage[T1]{fontenc}
\usepackage{tikz}
\usetikzlibrary{positioning,fit}
\DeclareMathAlphabet{\mathcal}{OMS}{cmsy}{m}{n}

\newcommand*{\balancecolsandclearpage}{%
  \close@column@grid
  \clearpage
  \twocolumngrid
}

\begin{document}

\title[Parrondo-type enhancement of quantum-state transfer in spin chains]{Parrondo-type enhancement of quantum-state transfer in spin chains}

\author[1]{\fnm{Rafael} \sur{Vieira}}\email{rafael.vieira1989@edu.udesc.br}
\equalcont{These authors contributed equally to this work.}

\author*[1]{\fnm{Edgard P. M.} \sur{Amorim}}\email{edgard.amorim@udesc.br}
\equalcont{These authors contributed equally to this work.}

\affil*[1]{\orgdiv{Departamento de F\'isica}, \orgname{Universidade do Estado de Santa Catarina}, \orgaddress{\postcode{89219-710}, \city{Joinville}, \state{SC}, \country{Brazil}}}

\abstract{Spin chains have been widely studied as quantum channels for short-distance communication in quantum devices, where many-body dynamics can mediate quantum-state transfer between distant sites. In finite unmodulated chains, however, dispersion and interference effects associated with the static Hamiltonian often limit the achievable transfer fidelity. Here we investigate the transfer of single-qubit and Bell states in finite $XX$ spin chains under periodic switching between two Hamiltonians with different boundary couplings. Inspired by Parrondo's paradox, we examine whether alternating between two configurations that individually yield suboptimal transfer fidelities can generate enhanced coherent transmission. Using Floquet theory together with numerical simulations in the single-excitation subspace, we show that periodic driving can outperform static configurations and achieve higher transfer fidelities. This enhancement originates from the noncommutativity of the driven Hamiltonians and reflects a purely coherent interference effect. We further analyze the dependence of the protocol on system size and driving parameters and examine its robustness to asymmetric boundary couplings. Our results show that the transfer fidelity remains stable under moderate disorder, indicating that simple time-dependent control of boundary couplings provides an effective strategy to enhance quantum-state transfer in spin-chain communication channels and optimize quantum information processing in engineered many-body systems.}



\maketitle

\section{Introduction}\label{sec:introduction}

The transfer of quantum states with high fidelity---that is, how closely the state received by a party matches the state prepared by the sender---is one of the central challenges in building scalable quantum devices \cite{Ben00}. As in the classical case, it is impossible to implement meaningful communication or distributed processing without a reliable mechanism to transmit quantum states between distant components of a quantum computer \cite{Ben84,Nie00}.

Quantum states can be transmitted between two parties, commonly referred to as Alice and Bob, through three conceptually distinct approaches. The first is the direct sending of the physical carrier that encodes the state, such as single photons propagating through optical fibers~\cite{Gis02}. Another approach is quantum teleportation, which relies on previously shared entangled qubit pairs and classical communication~\cite{Ben85}. A third option is to use many-body systems as quantum channels, in which the information propagates through the internal interactions of the system itself~\cite{Bos03}. Spin chains constitute a paradigmatic realization of this mechanism: Alice prepares a quantum state, which then evolves under the natural dynamics of the chain until it is retrieved by Bob at a later time. In this framework, the intermediate spins act as a quantum communication channel connecting the sender and the receiver.

Beyond their role as quantum communication channels, spin chains are also of considerable interest in condensed-matter physics. Recent studies have explored frustrated spin chains in correlated metals \cite{Moo25}, thermalization and localization phenomena in systems with single-ion anisotropy \cite{Sou24,Jaf25}, and the emergence of optimal quantum information scrambling in chiral chains exhibiting effective black-hole geometries \cite{Dan25}. Moreover, they have also been investigated as resources for quantum information processing tasks. For instance, several protocols exploit the transfer of quantum states through spin chains as a mechanism to implement quantum logic operations or distributed quantum gates \cite{Mar16}. These developments highlight the broad relevance of spin-chain dynamics, bridging fundamental questions in condensed-matter physics with practical challenges in quantum information processing.

In the simplest scenario of single-qubit transfer, Alice encodes a quantum state on a specific site of the chain, and the information propagates dynamically through the spin interactions until it reaches Bob at the opposite end of the channel \cite{Bos03,Chr04,Nik04,Sub04,Osb04,Chr05,Kar05,Har06,Huo08,Gua08,Ban10,Kur11,Apo12,Lor13,Kor14,Shi15,Zha15,Che16,Est17,Gra18,Beh18,Sha18,Bez25,Far25}. Numerous studies have shown that the natural dynamics generated by suitably engineered Hamiltonians can enable the transfer of quantum states with either high or perfect fidelity \cite{Bos03,Chr04,Chr05,Nik04,Osb04,Kar05}. Because of their simple structure and the possibility of controlling or engineering the local couplings, spin chains have therefore been extensively investigated as natural quantum channels for state transfer \cite{Bos03,Chr04,Chr05,Kar05,Har06,Ban10,Apo12,Vie18,Vie19,Vie20}. For the transfer of two non-entangled qubits, the dynamics of the chain must preserve the local correlations associated with each qubit during the propagation through the channel \cite{Bos03,Chr05,Li05,Har06,Ban10,Lor13,Est17,Apo10}. In contrast, the transfer of entangled two-qubit states poses an additional challenge, since the quantum correlations between the qubits must be preserved throughout the transport process. This requirement imposes more stringent constraints on the system Hamiltonian and on the structure of the spin-chain interactions \cite{Roj17,Est17a,Baz18,Vie18,Apo19,Vie19,Fre19,Vie20,Hos24,Str25}. 

However, the fidelity of state transfer in homogeneous spin chains is generally limited by dispersion, destructive interference, and sensitivity to imperfections in the system parameters~\cite{Sub04,Kar05,Gua08,Apo10,Ban10,Vie18,Vie19,Vie20}. Numerous schemes have been proposed to mitigate these effects, including engineered couplings~\cite{Chr04,Chr05,Kar05}, boundary control~\cite{Ban10,Apo12,Kor14}, and optimal control pulses~\cite{Shi15,Zha15}, but in many cases there remains a nontrivial trade-off between achievable fidelity, robustness to disorder, and protocol duration. This is particularly relevant for the transfer of entangled states (such as Bell states), where both the preservation of coherence and the correct distribution of correlations between distant sites are required~\cite{Vie18,Vie19,Vie20,Apo19}.

In this work we explore an alternative route based on Parrondo's paradox. Originally formulated in the context of classical game theory and inspired by Brownian ratchet models, Parrondo's effect describes the counterintuitive situation in which alternating between two individually losing games with an appropriate temporal pattern yields an overall winning strategy~\cite{parrondo1996criticism,harmer2001brownian,harmer2002review,amengual2004discrete,shu2014beyond}. This has been explored in a wide range of contexts, including biological adaptation~\cite{jean2021,cheong2019}, agricultural crop-rotation optimization~\cite{chaitanya2023}, the analysis of complex networks~\cite{cheong2016}, and the development of routing strategies~\cite{ankit2024}. In classical and quantum dynamical systems, this translates into the idea that the alternation between two unfavorable evolutions can give rise to an effective dynamics that is more advantageous than either of the static cases taken alone~\cite{lai2020review}. When applied to time-dependent Hamiltonians, a Parrondo-like strategy corresponds to switching between two distinct parameter regimes in such a way that their combined effect, over one or more periods of alternation, enhances a desired figure of merit~\cite{ximenes2024parrondo}.

Here we show that Parrondo's strategy can be used to enhance the fidelity of quantum-state transfer in spin-chain channels. Precisely, we consider a spin chain in which the couplings at the boundaries are not kept constant in time, but instead alternate periodically between two values: either two distinct values of the end coupling $\alpha$ or, alternatively, two values of the coupling $\beta$ that links the end blocks to the channel. Each static choice of coupling, when used alone, leads to suboptimal transfer of the initial state shared by Alice. Remarkably, by alternating these couplings in time with a fixed period, we demonstrate that the resulting stroboscopic dynamics can outperform both static scenarios, yielding a significantly higher transfer fidelity at specific times. 

From a quantum-information perspective, the goal is to preserve the coherence and entanglement encoded in the transmitted state while minimizing dispersion. In this context, the Parrondo protocol exploits coherent interference generated by periodically switching between two noncommuting Hamiltonians. The alternating dynamics reshapes the phases accumulated by the propagating excitation, modifying the interference pattern so that destructive interference is partially suppressed and constructive interference at the receiver is enhanced. Our results thus provide a concrete example of a Parrondo-like enhancement in a quantum state-transfer protocol and suggest that time-dependent control of boundary couplings can be an effective tool to improve entanglement distribution in spin-chain quantum channels.

This paper is organized as follows. In Sect.~\ref{sec:model} we introduce the spin-chain model, specify the static and time-dependent Hamiltonians that implement the Parrondo-type driving, and derive an effective Hamiltonian in the high-frequency limit. In Sect.~\ref{sec:fidelity} we define the figures of merit used throughout this work, namely the transfer fidelities for single-qubit and Bell states. In Sect.~\ref{sec:results} we analyze both transmission scenarios, investigating how the transfer fidelity in a modulated chain depends on the boundary couplings, identifying parameter regimes corresponding to losing strategies, and exploring different Parrondo-type protocols. In Sect.~\ref{sec:disorder} we study the robustness of these protocols against disorder. Finally, Sect.~\ref{sec:conclusions} summarizes the main results and outlines possible directions for future work.

\section{Mathematical model}\label{sec:model}

In this section, we introduce the spin-chain model and the Parrondo-type driving protocol used throughout this work. We first specify the static Hamiltonian with site-dependent couplings that defines the channel between Alice and Bob and describe how the boundary couplings are modulated in time, giving rise to a time-dependent Hamiltonian that implements the alternating (Parrondo) dynamics. Finally, we analyze the high-frequency limit of this protocol.

\subsection{Static and time-dependent Hamiltonians}\label{subsec:hamiltonians}

\begin{figure}[htp]
  \centering
  \resizebox{0.6\columnwidth}{!}{%
  \begin{tikzpicture}[
      every node/.style={font=\large},
      dot/.style={circle,minimum size=11mm,inner sep=0,
                  shading=ball,ball color=#1!60},
      link/.style={line width=2.5pt,draw=black},
      box/.style={rounded corners,draw=black,thick,inner sep=5pt}
    ]


    \node[dot=cyan] (A1a) at (0,0) {1};

    \node[dot=green, right=0.8cm of A1a] (C1a) {};      
    \node[dot=green, right=0.3cm of C1a] (C2a) {};      
    \node[dot=green, right=0.3cm of C2a] (C3a) {};      
    \node[right=0.01cm of C3a] (dotsA) {$\cdots$};
    \node[dot=green, right=0.01cm of dotsA] (C4a) {};   
    \node[dot=green, right=0.3cm of C4a] (C5a) {};      
    \node[dot=green, right=0.3cm of C5a] (C6a) {};      

    \node[dot=yellow, right=0.8cm of C6a] (BNa) {$N$};  

    \draw[link] (A1a) -- (C1a) node[midway,below=4pt] {$\alpha$};
    \draw[link] (C1a) -- (C2a) node[midway,below=4pt] {$\beta$};
    \draw[link] (C2a) -- (C3a) node[midway,below=4pt] {$\gamma$};
    \draw[link] (C4a) -- (C5a) node[midway,below=4pt] {$\gamma$};
    \draw[link] (C5a) -- (C6a) node[midway,below=4pt] {$\beta$};
    \draw[link] (C6a) -- (BNa) node[midway,below=4pt] {$\alpha$};

    \node[box,fit=(A1a),label={[yshift=1pt]Alice}] (AliceBoxA) {};
    \node[box,fit=(C1a)(C6a)(dotsA),label={[yshift=1pt]Channel}] (ChanBoxA) {};
    \node[box,fit=(BNa),label={[yshift=1pt]Bob}] (BobBoxA) {};
    \node[right=0.7cm of BobBoxA] {(a)};


    \begin{scope}[yshift=-2.6cm]

      \node[dot=cyan] (A1b) at (0,0) {1};
      \node[dot=cyan, right=0.3cm of A1b] (A2b) {2};

      \node[dot=green, right=0.8cm of A2b] (C1b) {};     
      \node[dot=green, right=0.3cm of C1b] (C2b) {};     
      \node[right=0.01cm of C2b] (dotsB) {$\cdots$};
      \node[dot=green, right=0.01cm of dotsB] (C3b) {};  
      \node[dot=green, right=0.3cm of C3b] (C4b) {};     

      \node[dot=yellow, right=0.8cm of C4b] (B1b) {$N\!-\!1$};
      \node[dot=yellow, right=0.3cm of B1b] (B2b) {$N$};

      \draw[link] (A1b) -- (A2b) node[midway,below=4pt] {$\alpha$};
      \draw[link] (A2b) -- (C1b) node[midway,below=4pt] {$\beta$};
      \draw[link] (C1b) -- (C2b) node[midway,below=4pt] {$\gamma$};
      \draw[link] (C3b) -- (C4b) node[midway,below=4pt] {$\gamma$};
      \draw[link] (C4b) -- (B1b) node[midway,below=4pt] {$\beta$};
      \draw[link] (B1b) -- (B2b) node[midway,below=4pt] {$\alpha$};

      \node[box,fit=(A1b)(A2b),label={[yshift=1pt]Alice}] (AliceBoxB) {};
      \node[box,fit=(C1b)(C4b)(dotsB),label={[yshift=1pt]Channel}] (ChanBoxB) {};
      \node[box,fit=(B1b)(B2b),label={[yshift=1pt]Bob}] (BobBoxB) {};
      \node[right=0.7cm of BobBoxB] {(b)};

    \end{scope}

  \end{tikzpicture}%
  }
  \caption{(Color online) Schematic spin chain with end blocks coupled to a uniform channel. \textbf{a} Alice transmits a single-qubit to Bob coupled to a channel with boundary couplings $\alpha$ and near-boundary couplings $\beta$. \textbf{b} Alice transmits a two-qubit state coupled by $\alpha$ to Bob with boundary couplings $\beta$. The spin channel couplings are $\gamma=1$}
  \label{fig:1}
\end{figure}
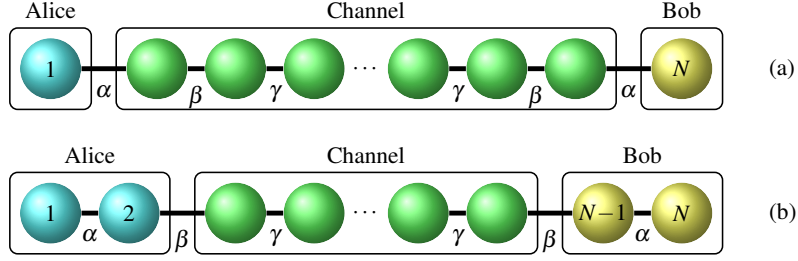

We consider a one-dimensional chain of $N$ spin-$1/2$ particles described by an $XX$ Hamiltonian with site-dependent nearest-neighbor couplings,  
\begin{equation}
    H = \sum_{i=1}^{N-1} J_{i,i+1}
    \bigl(\sigma_i^+ \sigma_{i+1}^- + \sigma_i^- \sigma_{i+1}^+\bigr),
    \label{eq:H-physical}
\end{equation}
where $\sigma_i^\pm$ are the spin raising and lowering operators acting on site $i$, and $J_{i,i+1}$ denotes the exchange coupling between sites $i$ and $i+1$. As shown schematically in Fig.~\ref{fig:1}, we identify the leftmost sites ($1$ and $2$) as Alice’s block and the rightmost sites ($N\!-\!1$ and $N$) as Bob’s block. These blocks are used to transmit (a) single-qubit and (b) two-qubit states between Alice and Bob. The intermediate sites between Alice and Bob blocks form a spin channel connecting them. The couplings are chosen to be symmetric with respect to the center of the chain,
\begin{equation}
    J_{1,2}=J_{N-1,N} = \alpha J_0, \quad J_{2,3}=J_{N-2,N-1} = \beta J_0, \quad J_{i,i+1}=\gamma J_0,\quad i = 3,4,\dots,N-3,
    \label{eq:couplings}
\end{equation}
where $J_0$ sets the overall energy scale and $\alpha,\beta,\gamma$ are dimensionless coupling parameters. For sake of simplicity, throughout this work we set $\gamma=1$.  

It is convenient to factor out $J_0$ and work with a dimensionless Hamiltonian, $H = J_0\,\mathcal{H}$, with
\begin{equation}
\mathcal{H}(\alpha,\beta;N)=
\alpha\left(\sigma_1^+\sigma_2^- + \sigma_{N-1}^+\sigma_N^- \right)
+\beta\left(\sigma_2^+\sigma_3^- + \sigma_{N-2}^+\sigma_{N-1}^- \right)
+\sum_{i=3}^{N-3}\sigma_i^+\sigma_{i+1}^-
+\mathrm{h.c.},
\label{eq:H-dimensionless-compact}
\end{equation}
Note that we should take $N\ge 6$ so that all three types of couplings in Eq.~\eqref{eq:couplings} are present. We define a dimensionless time variable
\begin{equation}
    \tau = \frac{J_0 t}{\hslash},
\end{equation}
so that $Ht/\hslash = \mathcal{H}\,\tau$ and the time-evolution operator for a time-independent Hamiltonian can be written as
\begin{equation}
    U(\tau) = e^{-i \mathcal{H}\,\tau}.
\end{equation}

\subsection{Periodic driving protocol}\label{subsec:driving}

To implement a Parrondo-like protocol, we consider a piecewise-constant time-dependent Hamiltonian in which one of the boundary couplings alternates between two values. In particular, we alternate the boundary coupling $\alpha$ or $\beta$ through a piecewise-constant periodic switching between two Hamiltonians,
\begin{equation}
\mathcal{H}(\tau)=
\begin{cases}
\mathcal{H}_1 = \mathcal{H}(\alpha_1,\beta_1,N),
& 0\le\tau<T_1,\\[6pt]
\mathcal{H}_2 = \mathcal{H}(\alpha_2,\beta_2,N),
& T_1\le\tau<T,
\end{cases}
\label{eq:H_piecewise}
\end{equation}
where the system evolves under $\mathcal{H}_1$ during a time interval $T_1$ and under $\mathcal{H}_2$ during a time interval $T_2=T-T_1$, yielding a total period $T=T_1+T_2$. We investigate two cases: (i) $\alpha_1\neq\alpha_2$ with $\beta_1=\beta_2$, and (ii) $\alpha_1=\alpha_2$ with $\beta_1\neq\beta_2$. We parametrize
\begin{equation}
T_1 = \frac{T}{2}+\delta T,
\qquad
T_2 = \frac{T}{2}-\delta T,
\label{T1T2}
\end{equation}
with $-\frac{T}{2}\le\delta T\le\frac{T}{2}$. Equivalently, defining $T=2\pi/\omega$, we write
\begin{equation}
T_1=\frac{2\pi}{\omega}\eta,
\qquad
T_2=\frac{2\pi}{\omega}(1-\eta),
\label{eq:periods}
\end{equation}
where
\begin{equation}
\eta=\frac{\delta T}{T}+\frac{1}{2},
\qquad
\eta\in[0,1].
\end{equation}
The time-evolution operator over a single period is then
\begin{equation}
U(T,0)=e^{-i\mathcal{H}_2 T_2}e^{-i\mathcal{H}_1 T_1}\equiv e^{-i\mathcal{H}_{\mathrm{eff}}T},
\label{eq:Floquet_def}
\end{equation}
and this pattern is repeated periodically with frequency $\omega$. The corresponding one-period evolution operator above defines an effective Floquet dynamics whose effective Hamiltonian $\mathcal{H}_{\mathrm{eff}}$ we further analyze. It is convenient to write 
\begin{equation}
e^{-i\mathcal{H}_{\mathrm{eff}}T}=e^{\Omega(T)},
\end{equation}
where $\Omega(T)$ is given by the Magnus expansion \cite{Mag54},
\begin{equation}
    \Omega(T) = \Omega_1(T) + \Omega_2(T) + \Omega_3(T) + \cdots,
\end{equation}
with the first terms
\begin{equation}
\Omega_1 = -i\int_0^T \mathcal{H}(\tau') d\tau',\quad
\Omega_2 = -\frac12 \int_0^T d\tau' 
\int_0^{\tau'} d\tau'' 
[\mathcal{H}(\tau'),\mathcal{H}(\tau'')],\quad
\Omega_3 \propto \mathcal{O}(T^3),
\end{equation}
We now evaluate $\Omega_1$ and $\Omega_2$ explicitly for the piecewise-constant Hamiltonian from Eq.~\eqref{eq:H_piecewise}. For the first term,
\begin{equation}
\Omega_1=
-i\left(
\mathcal{H}_1 T_1 + \mathcal{H}_2 T_2
\right)
=
-i\left[
\frac{\mathcal{H}_1+\mathcal{H}_2}{2}T
-
(\mathcal{H}_2-\mathcal{H}_1)\delta T
\right].
\label{eq:Omega1_final}
\end{equation}
Since $[\mathcal{H}_1,\mathcal{H}_1]=[\mathcal{H}_2,\mathcal{H}_2]=0$, only cross terms contribute to $\Omega_2$, giving
\begin{equation}
\Omega_2=-\frac12[\mathcal{H}_1,\mathcal{H}_2]
\int_{T_1}^{T} d\tau'
\int_0^{T_1} d\tau''
=
-\frac18
[\mathcal{H}_1,\mathcal{H}_2]
\left(
T^2 - 4\delta T^2
\right),
\label{eq:Omega2_final}
\end{equation}
where in the last equality we used $[\mathcal{H}_2,\mathcal{H}_1]=-[\mathcal{H}_1,\mathcal{H}_2]$.
Higher-order terms scale as $\Omega_n \propto T^n$ for $n\geq3$.

In the high-frequency regime $\omega\gg1$, i.e., $T\ll1$, we neglect terms of order $T^2$ and higher, obtaining
\begin{equation}
\mathcal{H}_{\mathrm{eff}}
=
\frac{\mathcal{H}_1+\mathcal{H}_2}{2}
-
\frac{\delta T}{T}
(\mathcal{H}_2-\mathcal{H}_1).
\label{eq:Heff_general}
\end{equation}
Since the $XX$ Hamiltonian depends linearly on the boundary couplings, the effective Hamiltonian can be written as
\begin{equation}
\mathcal{H}_{\mathrm{eff}}
=
\mathcal{H}
\left(
\alpha_{\mathrm{eff}},
\beta_{\mathrm{eff}};
N
\right),
\end{equation}
where
\begin{equation}
\alpha_{\mathrm{eff}}
=
\frac{\alpha_1+\alpha_2}{2}
-
\frac{\delta T}{T}
(\alpha_2-\alpha_1),\quad
\beta_{\mathrm{eff}}
=
\frac{\beta_1+\beta_2}{2}
-
\frac{\delta T}{T}
(\beta_2-\beta_1).
\end{equation}
For times that are integer multiples of the driving period, $\tau \approx mT$ with $m\in\mathbb{Z}$, the stroboscopic evolution can then be approximated as
\begin{equation}
U(\tau)
=
\left(
e^{-i\mathcal{H}_{\mathrm{eff}}T}
\right)^m
=
e^{-i\mathcal{H}_{\mathrm{eff}}\tau}.
\end{equation}

In the symmetric case $\delta T=0$, the effective Hamiltonian reduces to the arithmetic mean of $\mathcal{H}_1$ and $\mathcal{H}_2$. For finite $\delta T$, asymmetry between time intervals produces corrections that may enhance state-transfer performance. Away from the strict high-frequency limit, the nonvanishing commutator $[\mathcal{H}_1,\mathcal{H}_2]$ generates higher-order Floquet corrections, making the ordering of the Hamiltonians physically relevant. As a consequence, for low frequencies, the order dependence of the driven Hamiltonians leads to a coherent interference effect and the enhancement of fidelity.

This interpretation relies on the effective Hamiltonian obtained from the Magnus expansion, whose validity is restricted to the high-frequency regime and becomes progressively less accurate at moderate driving frequencies. As shown by Fel'dman \cite{Fel84}, the Magnus expansion for periodically driven systems has a limited convergence domain. Crucially, the largest fidelity enhancements in our Parrondo protocol occur in a regime where the perturbative approximation is close to the edge of its validity, suggesting that the improvement mechanism is inherently nonperturbative. The numerical results presented here are computed from the exact time evolution and therefore do not rely on the validity of the effective-Hamiltonian approximation.

The validity of the effective-Hamiltonian description can be assessed by comparing the fidelity obtained from
the exact periodically driven evolution with the fidelity predicted by the high-frequency effective Hamiltonian
$\mathcal{H}_{\mathrm{eff}}$. In the high-frequency regime, the exact driven dynamics approaches the effective-Hamiltonian prediction. At lower driving frequencies, however, visible deviations arise due to higher-order Floquet corrections generated by the noncommutativity of $\mathcal{H}_1$ and $\mathcal{H}_2$. These deviations are precisely the finite-frequency effects responsible for the Parrondo enhancement observed in the protocol.

Unlike optimization-based approaches that rely on continuous searches in control space \cite{Pec23a,Pec23b}, our Parrondo-
inspired protocol enhances quantum-state transfer through noncommutative switching between discrete cou-
pling configurations. Similarly, although time-dependent control localized near the receiver can also improve
transfer in spin chains \cite{Pec24}, such protocols differ from the present mechanism, since they rely on magnetic-field
or coupling reshaping rather than periodic switching between two individually suboptimal Hamiltonians.

\section{Fidelity of quantum states}\label{sec:fidelity}

Because the XX Hamiltonian conserves the total number of excitations, the dynamics induced by $U(\tau)$ is block diagonal in excitation-number sectors. Since the protocol is based on deterministic unitary evolution without measurements or post-selection, there is no independent notion of success probability. Instead, its performance is characterized by the state-transfer fidelity. For a pure single-excitation input, this fidelity is equal to the probability of finding the excitation at the receiving site. In this work we restrict to the vacuum and single-excitation subspaces. A general state in this sector can be written as
\begin{equation}
\ket{\psi(\tau)} = a_0(\tau)\ket{\mathbf{0}} + \sum_{k=1}^{N} a_k(\tau)\ket{\mathbf{k}},
\label{eq:psi_general}
\end{equation}
where $\ket{\mathbf{0}}=\ket{0_1 0_2\cdots 0_N}$ is the vacuum (no excitation) and
$\ket{\mathbf{k}}=\ket{0_1\cdots 0_{k-1}1_k0_{k+1}\cdots 0_N}$ denotes a single excitation localized at site $k$. Normalization implies $|a_0(\tau)|^2+\sum_{k=1}^{N}|a_k(\tau)|^2=1$. 

We distinguish between (i) single-qubit state transfer and (ii) bipartite (Bell-state) transfer as follows.

\subsection{Single-qubit state transfer}\label{subsec:single_qubit}

We first consider an input qubit state encoded at site $1$ in the standard Bloch-sphere parametrization,
\begin{equation}
\ket{\psi(0)}=
\cos\!\left(\frac{\theta}{2}\right)\ket{\mathbf{0}}
+ e^{i\phi}\sin\!\left(\frac{\theta}{2}\right)\ket{\mathbf{1}},
\label{eq:psi_single_in}
\end{equation}
where $\theta\in[0,\pi]$ and $\phi\in[0,2\pi)$ are the polar and azimuthal angles.
After time $\tau$, the state becomes
\begin{equation}
\ket{\psi(\tau)}
=U(\tau)\ket{\psi(0)}
=
\cos\!\left(\frac{\theta}{2}\right)\ket{\mathbf{0}}
+ e^{i\phi}\sin\!\left(\frac{\theta}{2}\right)
\sum_{k=1}^{N} f_{k,1}(\tau)\ket{\mathbf{k}},
\label{eq:psi_single_evol}
\end{equation}
where
\begin{equation}
f_{k,1}(\tau)\equiv \braket{\mathbf{k}|U(\tau)|\mathbf{1}}
\label{eq:fk1_def}
\end{equation}
is the transition amplitude from site $1$ to site $k$ at time $\tau$. Therefore,
\begin{equation}
a_0(\tau)=\cos\!\left(\frac{\theta}{2}\right),\qquad
a_k(\tau)=e^{i\phi}\sin\!\left(\frac{\theta}{2}\right)f_{k,1}(\tau).
\label{eq:ak_single}
\end{equation}

To assess the received single-qubit state at Bob's site $N$, we compute the reduced density matrix
$\rho_N(\tau)=\text{Tr}_{\overline{N}}\!\left[\rho(\tau)\right]$, where $\rho(\tau)=\ket{\psi(\tau)}\bra{\psi(\tau)}$ and the trace is taken over all sites except $N$. The reduced density matrices are obtained by tracing out the degrees of freedom that do not belong to Bob's subsystem. Explicitly,
\begin{equation}
\rho_{N}(\tau)
=
\mathrm{Tr}_{1,\ldots,N-1}
\left[
|\psi(\tau)\rangle\langle\psi(\tau)|
\right]
=
\sum_{i_1,\ldots,i_{N-1}}
\langle i_1\cdots i_{N-1}|\psi(\tau)\rangle
\langle\psi(\tau)|i_1\cdots i_{N-1}\rangle,
\label{rhoN}
\end{equation}
when ${\ket{i_k}}$ is the computacional basis of the $N-1$ qubits. In the local basis of Bob's qubits $\{\ket{0}_N,\ket{1}_N\}$, one obtains
\begin{equation}
\rho_N(\tau)\!=\!
\begin{bmatrix}
1\!-\!\sin^2\!\left(\frac{\theta}{2}\right)\!|{f_{N,1}(\tau)}|^2
&
e^{-i\phi}\frac{\sin\theta}{2}f_{N,1}^*(\tau)
\\[4pt]
e^{i\phi}\frac{\sin\theta}{2}f_{N,1}(\tau)
&
\sin^2\!\left(\frac{\theta}{2}\right)\!|f_{N,1}(\tau)|^2
\end{bmatrix}\!.
\label{eq:rhoN}
\end{equation}
Up to an irrelevant global phase, the target output qubit at site $N$ should be
\begin{equation}
\ket{\psi_N}=
\cos\!\left(\frac{\theta}{2}\right)\ket{0}_N
+
e^{i\phi}\sin\!\left(\frac{\theta}{2}\right)\ket{1}_N.
\label{eq:psi_out_single}
\end{equation}
We define the state-transfer fidelity as
\begin{equation}
F(\tau)=\bra{\psi_N}\rho_N(\tau)\ket{\psi_N},
\label{eq:F_single_def}
\end{equation}
which yields
\begin{equation}
    F(\tau)=\cos^2\!\left(\frac{\theta}{2}\right)+\frac{\sin^2\theta}{2}\,\Re\!\left[f_{N,1}(\tau)\right]-\cos\theta\,\sin^2\!\left(\frac{\theta}{2}\right)|f_{N,1}(\tau)|^2,
\label{eq:F_single}
\end{equation}
being independent of the phase $\phi$, as expected from excitation-number conservation: the initial phase $e^{i\phi}$ multiplies the entire single-excitation component and cancels in the overlap defining the fidelity. In the particular case $\theta=\pi$ corresponding to the input of pure excitation, $\ket{\psi(0)}=\ket{1}$, the fidelity reduces to $F(\tau)=|f_{N,1}(\tau)|^2=|a_N(\tau)|^2$.

\subsection{Bell-state transfer}\label{subsec:bell}

We initialize the chain with the following Bell states, $\ket{\psi^\pm_{1,2}}=\left(\ket{01}_{1,2}\pm\ket{10}_{1,2}\right)/\sqrt{2}$
localized in Alice's block on sites $(1,2)$,
\begin{equation}
\ket{\psi^\pm(0)}=\frac{1}{\sqrt{2}}\left(\ket{\mathbf{1}}\pm\ket{\mathbf{2}}\right),
\label{eq:bell_in}
\end{equation}
which lies entirely in the single-excitation subspace (hence $a_0(\tau)=0$ for all $\tau$).

To quantify the transfer to Bob's sites $(N-1,N)$, we compute the reduced density matrix tracing out the first $N-2$ spins, $\rho_{N-1,N}(\tau)=\text{Tr}_{\overline{N-1,N}}\!\left[\rho(\tau)\right]$. As in Eq.(\ref{eq:rhoN}), the partial trace over the degrees of freedom that do not belong to Bob's subsystem is given by
\begin{equation}
\rho_{N-1,N}(\tau)
=
\mathrm{Tr}_{1,\ldots,N-2}
\left[
|\psi(\tau)\rangle\langle\psi(\tau)|
\right]
=\!\!\!\!
\sum_{i_1,\ldots,i_{N-2}}\!\!\!\!\!
\langle i_1\cdots i_{N-2}|\psi(\tau)\rangle
\langle\psi(\tau)|i_1\cdots i_{N-2}\rangle .
\end{equation}

The reduced density matrix for Bob's qubits on the two-qubit basis $\{\ket{00},\ket{01},\ket{10},\ket{11}\}$, whose ordering refers to sites $(N-1,N)$, then reads
\begin{equation}
    \rho_{N-1,N}(\tau)=
    \begin{bmatrix}
    1-|a_{N-1}(\tau)|^2-|a_{N}(\tau)|^2 & 0 & 0 & 0 \\
    0 & |a_{N}(\tau)|^2 & a_{N}(\tau) a_{N-1}^*(\tau) & 0 \\
    0 & a_{N-1}(\tau) a_{N}^*(\tau) & |a_{N-1}(\tau)|^2 & 0 \\
    0 & 0 & 0 & 0
    \end{bmatrix}.
    \label{eq:rhoab}
\end{equation}
The target output Bell state on Bob's sites is
\begin{equation}
\ket{\psi^\pm_{N-1,N}}
=\frac{1}{\sqrt{2}}\left(\ket{01}_{N-1,N}\pm\ket{10}_{N-1,N}\right),
\label{eq:bell_target}
\end{equation}
and we define the Bell-state transfer fidelity as
\begin{equation}
F(\tau)
=\bra{\psi^\pm_{N-1,N}}\rho_{N-1,N}(\tau)\ket{\psi^\pm_{N-1,N}}.
\label{eq:F_bell_def}
\end{equation}
Using Eq.~\eqref{eq:rhoab}, we obtain
\begin{equation}
F(\tau)=\frac{1}{2}\,|a_{N-1}(\tau)\pm a_N(\tau)|^2.
\label{eq:F_bell}
\end{equation}
Therefore, perfect transfer of the initial Bell state from Alice to Bob, $F(\tau)=1$, is achieved whenever the amplitudes at the last two sites satisfy $a_{N-1}(\tau) = \pm a_N(\tau)=\pm 1/\sqrt{2}$. 

For completeness, we also evaluate the concurrence of the reduced two-qubit state at Bob’s sites \cite{Hill97,Wootter98}. This
quantity provides a direct measure of the entanglement shared between the receiver qubits and is used as a
complementary indicator of transfer quality,
\begin{equation}
    C(\rho_{N-1,N})=\max\left\{0,\lambda_1-\lambda_2-\lambda_3-\lambda_4\right\},
\end{equation}
where $\lambda_{1,2,3,4}$  are the square roots, in decreasing order, of the eigenvalues of $R=\rho_{N-1,N} (\sigma_y \otimes \sigma_y) \rho_{N-1,N}^*(\sigma_y \otimes \sigma_y)$.

\section{Results} \label{sec:results}

Our goal is to investigate whether periodic modulation of the spin-chain couplings can enhance the fidelity of quantum-state transmission between Alice and Bob beyond the performance achieved by static configurations. In particular, we analyze whether alternating in time between Hamiltonians that individually yield lower transmission fidelities can, under suitable driving frequencies, generate a higher overall fidelity. To this end, we compare the periodically driven protocol and static configurations with the benchmark uniform chain, where all couplings are equal ($\alpha=\beta=1$).

We therefore search for regimes in which two static Hamiltonians $\mathcal{H}_1$ and $\mathcal{H}_2$, each yielding fidelities lower than that of the uniform chain, can produce a higher transmission fidelity when alternated periodically. More precisely, let $F_{\mathcal{H}_1}$ and $F_{\mathcal{H}_2}$ denote the fidelities obtained from the evolution under the static Hamiltonians $\mathcal{H}_1$ and $\mathcal{H}_2$, $F_0$ the fidelity of the uniform chain, and $F_P$ the fidelity obtained under periodic modulation between $\mathcal{H}_1$ and $\mathcal{H}_2$. When the following conditions are simultaneously satisfied,
\begin{equation}
F_{\mathcal{H}_1}< F_0,\qquad F_{\mathcal{H}_2}< F_0,\qquad F_P> F_0,
\end{equation}
we obtain a Parrondo-like enhancement, defining a quantum analogue of the Parrondo effect in the context of quantum state transfer.

In our analysis, we focus on the first arrival peak of the fidelity curve, corresponding to the earliest time $\tau$ at which the transmitted state reaches Bob's block. This peak represents the shortest transmission time with maximal fidelity before significant dispersion occurs. For each set of driving parameters $(\omega,\eta,\alpha,\beta)$, the first-arrival time is obtained numerically by locating the earliest local maximum of the fidelity. Consequently, the measurement time is uniquely determined by the selected protocol parameters and does not constitute an additional optimization variable. In an experimental implementation, this time can be calibrated once for a given parameter set through preliminary characterization of the transfer dynamics and then used for subsequent repetitions of the protocol.

For simplicity, we restrict the control parameters to the couplings that connect the boundary qubits (Alice and Bob) to the transmission channel. Consequently, we consider two different scenarios: (i) the transfer of single-qubit states, in which we vary only the boundary coupling $\alpha$ while keeping the second-neighbor coupling fixed at $\beta=1$; and (ii) the transfer of Bell states, in which we keep $\alpha=1$ fixed and vary the parameter $\beta$, which controls the coupling between the boundary region and the bulk of the chain. 

\subsection{Single-qubit transfer}\label{subsec:single_particle}

We begin by analyzing the transfer of a single-excitation state. As a first step, we evaluate the transmission fidelity for static values of the coupling parameter $\alpha$, in order to identify which values lead to a losing evolution, i.e., with fidelity lower than that obtained in the uniform case $\alpha_0=1$, and which correspond to a winning evolution. In this analysis we restrict the parameter range to couplings of the same order of magnitude as the bulk interactions, since it is well known that in the regimes $\alpha\ll1$ or $\beta\ll1$ nearly perfect state transfer can be achieved after sufficiently long evolution times $\tau$ \cite{Vie18,Apo19,Vie19,Vie20}.

\begin{figure}[htp]
\centering
\includegraphics[width=0.4\linewidth]{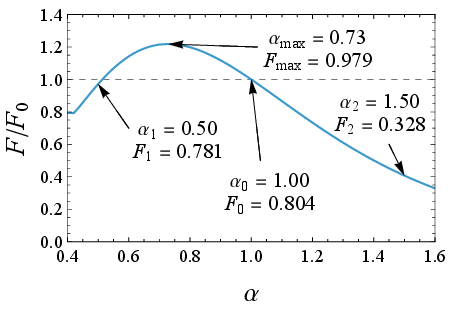}
\caption{Ratio $F/F_0$ obtained at the first arrival transmission peak when Alice sends the state $\ket{1}$ to Bob through a chain with $N=10$ qubits. Here $F$ denotes the maximum fidelity for a given coupling constant $\alpha$, while $F_0$ corresponds to the reference case $\alpha_0=1$, for which $F_0=0.804$. The dashed line indicates the boundary between winning (above the line) and losing (below the line) transmission evolutions. The markers highlight two losing examples given by $\alpha_1=0.50$ and $\alpha_2=1.50$, and a winning case with $\alpha_{\mathrm{max}}=0.73$, which yields the maximum fidelity within this parameter range}
\label{Fig2}
\end{figure}

To define the transmission protocol, we first specify the initial state prepared by Alice. From Eq.~\eqref{eq:psi_single_evol}, we choose $\theta=\pi$ and $\phi=0$, which yield the initial state $\ket{\psi(0)}=\ket{1}$. This choice is motivated by the fact that, according to Eq.~\eqref{eq:F_single}, only the excited component effectively participates in the dynamics. Consequently, demonstrating efficient transfer for this state suffices to characterize the behavior of more general input states. Figure~\ref{Fig2} shows the ratio between the maximum fidelity at the first arrival peak at Bob, $F(\alpha)$, and the reference value $F_0=0.804$, corresponding to the uniform chain ($\alpha_0=1$). The line $F/F_0=1$ marks the boundary separating winning ($F>F_0$) and losing ($F<F_0$) configurations.

Within this framework, two possible routes for observing the Parrondo effect emerge. The first corresponds to the analytical solution obtained in the high-frequency limit discussed in Sect.~\ref{subsec:driving}. In this regime, it is possible to select two losing values of $\alpha$, denoted $\alpha_1$ and $\alpha_2$, together with a relative period deviation $\eta$, such that the effective parameter $\alpha_{\mathrm{eff}}=\langle\alpha\rangle-\eta\Delta\alpha$ coincides with a value that produces a winning evolution, i.e., one whose transmission fidelity exceeds that of the uniform chain. The second possibility, which we explore here, arises when the driving frequency is not sufficiently large for the high-frequency approximation to hold. 

The Magnus expansion serves only as an analytical tool for understanding the high-frequency limit of the periodically driven system. In particular, it provides an effective Floquet Hamiltonian that clarifies how the alternating protocol reduces, at high frequency, to an effective static Hamiltonian with renormalized boundary couplings. However, all numerical fidelities reported in this work were computed from the exact unitary evolution generated by the piecewise-constant Hamiltonian in Eq.~(\ref{eq:H_piecewise}), implemented in the single-excitation subspace. Therefore, the reported Parrondo enhancements do not rely on the lowest-order Magnus approximation. Indeed, several of the largest fidelity enhancements occur at finite driving frequencies, where deviations from the high-frequency approximation become relevant.

\begin{figure}[htp]
\centering
\includegraphics[width=0.4\linewidth]{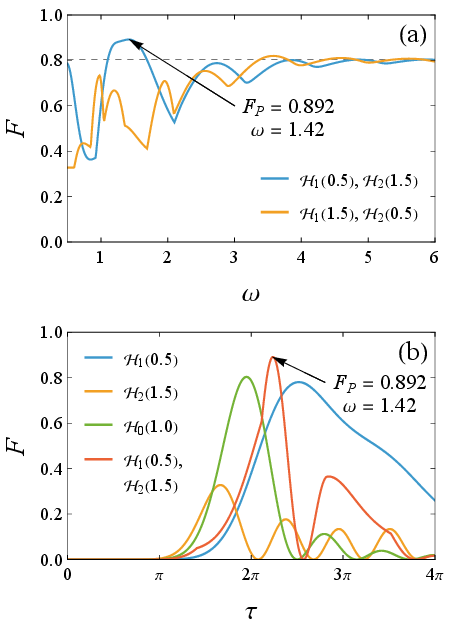}
\caption{\textbf{a} Maximum fidelity $F$ at the first transmission peak when Alice sends the state $\ket{1}$ from site $1$ to Bob at site $N$, as a function of the Parrondo driving frequency $\omega$ for a chain with $N=10$ qubits. The blue and orange curves correspond to evolutions starting with the losing Hamiltonians $\mathcal{H}_1(0.5)$ and $\mathcal{H}_2(1.5)$, respectively. The dashed line indicates the maximum fidelity obtained in the static uniform case $\alpha_0=1$. As $\omega$ increases,
both curves approach the high-frequency limit corresponding to the effective uniform coupling $\alpha_{eff}=1$. \textbf{b} Comparison of the transmission fidelity for the individual Hamiltonians and the periodically driven protocol, showing that the Parrondo protocol (red curve) outperforms both static losing strategies}
\label{Fig3}
\end{figure}

The losing parameters were chosen on opposite sides of the optimal static coupling because preliminary scans indicate that selecting both parameters in the region $\alpha>1$ leads to significantly weaker enhancements, reflecting the monotonic decrease of the static fidelity away from the optimum. For the numerical analysis we consider a chain of size $N=10$ and choose $\alpha_1=0.5$ and $\alpha_2=1.5$, two values corresponding to losing static configurations (see Fig.~\ref{Fig2}). We initially set $\eta=0$, which implies $\alpha_{\mathrm{eff}}=1$ in the asymptotic high-frequency limit. Figure~\ref{Fig3}a shows a scan over driving frequencies outside the asymptotic regime. We observe a frequency at which the transmission fidelity becomes significantly higher than in the uniform case when the evolution starts with $\mathcal{H}_1(0.5)$ and the Hamiltonians are alternated with frequency $\omega/2=0.71$. This result demonstrates that periodically alternating two losing strategies can indeed produce an effective winning dynamics, which is the defining feature of the Parrondo effect. Figure~\ref{Fig3}b compare the situation in which the average of the two losing parameters $\alpha_1$ and $\alpha_2$  coincides with the uniform value, thereby highlighting that certain driving frequencies $\omega$ can yield fidelities exceeding the asymptotic limit of the reference regime. We then explore more favorable scenarios by introducing asymmetry between the evolution times associated with each Hamiltonian while keeping the total driving period fixed.

Lastly, we incorporate the parameter $\eta$ into the analysis in order to investigate how the asymmetry between the evolution periods $T_1$ and $T_2$ affects the efficiency of quantum-state transmission from Alice to Bob. In this study, we select pairs of losing parameters $\alpha_1$ and $\alpha_2$ chosen as the closest values to $\alpha=1$, located respectively below and above the reference regime. Table~\ref{Tab1} presents a systematic parameter sweep in which the chain length is varied in the range $8 \le N \le 20$, while scanning the parameter space ($\omega$, $\eta$). For each configuration we identify the parameters that maximize the transmission fidelity under the periodically driven protocol. The results show that the maximum fidelity remains above $99\%$ for chain lengths up to $N=14$. Even for longer chains, the protocol continues to provide a substantial improvement. For instance, for $N=20$ the periodically driven protocol with unequal evolution periods increases the fidelity from $F_0=0.632$ in the uniform regime to $F_P=0.973$. These results demonstrate that temporal asymmetry in the driving protocol provides an efficient mechanism for maintaining high-fidelity state transfer even as the chain length increases.
\begin{table}[htp]
\centering
\caption{Optimal parameters for the Parrondo protocol obtained by periodically alternating the Hamiltonians $\mathcal{H}_1(\alpha_1)$ and $\mathcal{H}_2(\alpha_2)$. The parameters $\alpha_1$ and $\alpha_2$ correspond to individually losing static configurations located below and above the reference value $\alpha=1$, respectively. In all simulations $\mathcal{H}_1(\alpha_1)$ is applied first in the evolution sequence. For each chain length $N$, $F_0$ was determined and a parameter sweep over $0.50\le\omega\le3.50$ and $0.00\le\eta\le1.00$ was performed to determine the configuration that maximizes the transmission fidelity. The last column reports the maximum fidelity obtained with the Parrondo protocol}
\label{Tab1}
\begin{tabular*}{\columnwidth}{@{\extracolsep{\fill}}ccccccccc}
\hline
$N$ & $F_0$ & $\alpha_1$ & $F_{\mathcal{H}_1}(\alpha_1)$ & $\alpha_2$ & $F_{\mathcal{H}_2}(\alpha_2)$ & $\omega$ & $\eta$ & $F_P$ \\
\hline
8  & 0.854 & 0.55 & 0.843 & 1.01 & 0.845 & 1.46 & 0.51 & 0.997 \\
9  & 0.828 & 0.53 & 0.822 & 1.01 & 0.818 & 1.29 & 0.46 & 0.993 \\
10 & 0.804 & 0.51 & 0.798 & 1.01 & 0.793 & 1.14 & 0.42 & 0.989 \\
11 & 0.781 & 0.49 & 0.771 & 1.01 & 0.770 & 2.01 & 0.65 & 0.991 \\
12 & 0.760 & 0.47 & 0.742 & 1.01 & 0.749 & 1.84 & 0.64 & 0.993 \\
13 & 0.740 & 0.46 & 0.729 & 1.01 & 0.729 & 1.71 & 0.64 & 0.994 \\
14 & 0.722 & 0.45 & 0.716 & 1.01 & 0.710 & 1.60 & 0.63 & 0.990 \\
15 & 0.705 & 0.44 & 0.702 & 1.01 & 0.693 & 1.50 & 0.62 & 0.984 \\
16 & 0.688 & 0.43 & 0.687 & 1.01 & 0.676 & 2.08 & 0.65 & 0.975 \\
17 & 0.673 & 0.42 & 0.671 & 1.01 & 0.661 & 1.95 & 0.66 & 0.977 \\
18 & 0.659 & 0.41 & 0.655 & 1.01 & 0.646 & 1.85 & 0.66 & 0.977 \\
19 & 0.645 & 0.40 & 0.638 & 1.01 & 0.633 & 1.75 & 0.67 & 0.976 \\
20 & 0.632 & 0.39 & 0.626 & 1.01 & 0.620 & 1.66 & 0.67 & 0.973 \\
\hline
\end{tabular*}
\end{table}

\subsection{Bell-state transfer}

We examine the transfer of a Bell state given by Eq.~\eqref{eq:bell_in}, initially prepared on Alice's qubits $(1,2)$ by investigating the effect of varying the parameter $\beta$. In analogy with the analysis performed for variations in $\alpha$, we restrict the study to values close to the reference case $\beta=1$. Throughout this analysis the coupling between these qubits is kept fixed at $\alpha=1$, while Bob accesses the qubits $(N-1,N)$ at the opposite end of the chain, whose internal coupling is also fixed. As in the previous section, we focus on the fidelity associated with the first transmission peak. This choice is motivated by the fact that even small accumulated disorder in the system renders subsequent peaks less predictable, whereas the first peak remains robust and therefore provides a reliable measure of transmission performance.

\begin{figure}[htp]
\centering
\includegraphics[width=0.4\linewidth]{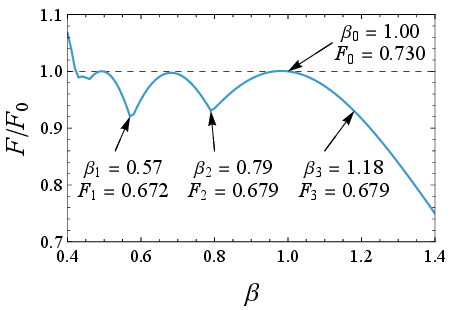}
\caption{Ratio $F/F_0$ at the first transmission peak obtained when Alice sends the Bell state $\ket{\psi^\pm}$ to Bob as a function of the coupling parameter $\beta$ for a chain with $N=10$ qubits. Here $F$ denotes the maximum fidelity for a given $\beta$, while $F_0=0.730$ corresponds to the reference value obtained for $\beta_0=1$. For $\beta<0.40$ the system enters a regime of almost perfect transmission earlier investigated \cite{Apo19}, which is not considered here since the objective is to improve transmission using individually losing parameters as $\beta_1$, $\beta_2$, and $\beta_3$ above}
\label{Fig4}
\end{figure}

For a chain of size $N=10$, we first identify values of $\beta$ for which the maximum fidelity at the first peak satisfies $F/F_0<1$, where $F_0$ denotes the fidelity obtained in the reference case $\beta=1$. Figure~\ref{Fig4} shows the ratio $F/F_0$ as a function of $\beta$. Three representative losing values of $\beta$ are selected and used as candidates for the Parrondo protocol.

Similarly to the study of variations in $\alpha$, we search for driving frequencies $\omega$ that maximize the transmission fidelity at the first peak. Table~\ref{tab3} summarizes the results obtained by considering all pairwise combinations of the three selected values of $\beta$. For each pair we also investigate the dependence on the order in which the parameters are applied, identifying which value initiates the evolution ($\beta_1$) and which acts as the alternating parameter ($\beta_2$). When each parameter is considered individually (i.e., without alternation), the maximum fidelities obtained satisfy $F<0.680$, indicating losing static configurations. However, when the parameters are periodically alternated with a suitable driving frequency $\omega$, the resulting fidelities $F_P$ increase significantly, ranging between $0.773$ and $0.836$. All these values exceed the fidelity obtained in the reference case $\beta_0=1$, where $F_0=0.730$.

An important feature revealed by Table~\ref{tab3} is that the order of application of $\beta_1$ and $\beta_2$ is not symmetric. Exchanging the parameters modifies both the optimal driving frequency $\omega$ and the final fidelity. In particular, the highest fidelity, $F_P=0.836$, is obtained when the evolution starts with $\beta_1=0.57$ and alternates with $\beta_2=0.79$. This result shows that the optimal combinations do not necessarily involve values below and above $\beta_0=1$ simultaneously; rather, the improvement emerges from an effective parameter $\beta_{\mathrm{eff}}$ that lies near the optimal region predicted by the analytical analysis.

\begin{table}[htp]
\centering
\caption{Optimal transmission fidelities obtained using the Parrondo protocol with $N=10$ for different combinations of the coupling parameters $\beta_1$ and $\beta_2$. In all cases the resulting fidelities exceed the reference value $F_0=0.730$ obtained for $\beta_0=1$. The table also illustrates three scenarios highlighting the dependence of the protocol on the order in which the parameters ($\beta_1$, $\beta_2$) are applied}
\label{tab3}

\begin{tabular*}{\columnwidth}{@{\extracolsep{\fill}}ccccc}
\hline
$\beta_1$ & $\beta_2$ & $\omega$ & $\eta$ & $F_P$ \\
\hline
0.57 & 0.79 & 0.95 & 0.48 & 0.836 \\
0.57 & 1.18 & 1.05 & 0.66 & 0.873 \\
0.79 & 0.57 & 0.57 & 0.59 & 0.773 \\
0.79 & 1.18 & 2.17 & 0.50 & 0.791 \\
1.18 & 0.57 & 1.57 & 0.61 & 0.820 \\
1.18 & 0.79 & 1.52 & 0.54 & 0.792 \\
\hline
\end{tabular*}
\end{table}

\begin{figure}[htp]
\centering
\includegraphics[width=0.4\linewidth]{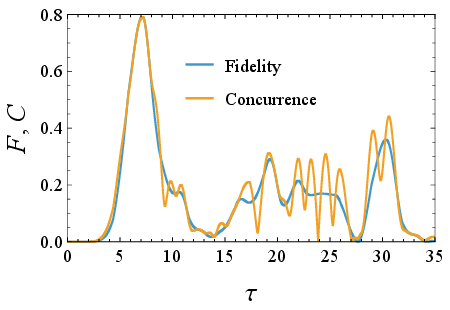}
\caption{Comparison between the Bell-state transfer fidelity and concurrence for a chain with $N=10$. The driving parameters are fixed at $\alpha=1$, $\gamma=1$, $\beta_1=0.79$, $\beta_2=1.18$, $\omega=2.17$, and $\eta=0.50$. The concurrence quantifies the amount of entanglement present in Bob's reduced state, whereas the fidelity measures the overlap with the specific Bell state initially prepared by Alice. The absence of a large concurrence before the first fidelity maximum supports the use of the first fidelity peak as the relevant transfer time}
\label{Fig5}
\end{figure}

At this point, a natural question is whether highly entangled states other than the one prepared by Alice can reach Bob before the first fidelity maximum. To address this point, Fig.~\ref{Fig5} compares the fidelity and concurrence for a chain with $N=10$. While concurrence quantifies the amount of entanglement present in Bob's reduced state, it does not measure its overlap with the specific state initially prepared by Alice. The comparison shows that no highly entangled state reaches Bob before the first fidelity maximum when Alice prepares, for example, the initial state $\ket{\Psi^{+}}$. Therefore, the first fidelity peak identifies not only the earliest arrival of a strongly entangled state, but also the earliest time at which the transmitted state faithfully reproduces the quantum state prepared by Alice.

We next extend the analysis to chains with different lengths $N$. For each size we select two losing parameters, $\beta_1<1$ and $\beta_2>1$, such that both individually yield fidelities lower than the reference value when applied independently. The results are shown in Table~\ref{tab2} for chain sizes $8\le N\le12$. In all cases, periodic alternation between these parameters produces measurable gains in the transmission fidelity relative to the reference case $\beta_0=1$. Although the improvement decreases as the system size increases, the Parrondo protocol remains advantageous for all considered chain lengths. These results indicate that the Parrondo effect persists even as the system size grows, acting as an effective mechanism for mitigating fidelity loss in short spin chains.

\begin{table}[htp]
\centering
\caption{Transmission fidelities obtained for the Parrondo protocol applied to Bell-state transfer in chains with $8\le N\le12$. For each system size, two losing parameters $\beta_1<1$ and $\beta_2>1$ are selected close to the reference value $\beta_0=1$. The parameter space $0.00\le\omega\le3.00$ and $0.00\le\eta\le1.00$ is scanned to determine the optimal transmission fidelity}
\label{tab2}

\begin{tabular*}{\columnwidth}{@{\extracolsep{\fill}}ccccccccc}
\hline
$N$ & $F_0$ & $\beta_1$ & $F_{\mathcal{H}_1}(\beta_1)$ & $\beta_2$ & $F_{\mathcal{H}_2}(\beta_2)$ & $\omega$ & $\eta$ & $F_P$ \\
\hline
8  & 0.807 & 0.77 & 0.717 & 1.28 & 0.716 & 0.98 & 0.22 & 0.962 \\
9  & 0.767 & 0.78 & 0.694 & 1.23 & 0.694 & 0.88 & 0.03 & 0.879 \\
10 & 0.730 & 0.79 & 0.679 & 1.18 & 0.679 & 2.17 & 0.50 & 0.791 \\
11 & 0.696 & 0.81 & 0.662 & 1.14 & 0.661 & 1.99 & 0.50 & 0.746 \\
12 & 0.664 & 0.80 & 0.653 & 1.06 & 0.654 & 1.78 & 0.43 & 0.706 \\
\hline
\end{tabular*}
\end{table}

Even in situations where individually losing parameters are not available, periodic Hamiltonian alternation within the Floquet framework can still act as a fidelity optimization mechanism. In particular, if optimal values exist between $\beta_1$ and $\beta_2$, the maximum achievable fidelity approaches that obtained for the effective coupling $\beta_{\mathrm{eff}}$ in the high-frequency limit.

\section{Disorder analysis}\label{sec:disorder}

In order to assess the robustness of the proposed protocol, we analyze how deviations from the ideal symmetric coupling configuration and temporal noise affect the transmission fidelity. 

\subsection{Asymmetries between the coupling parameters}
In particular, we consider asymmetries between the coupling parameters on Alice's and Bob's sides of the spin chain. While the couplings on Alice's side are kept at $J_{1,2}/J_0=\alpha$ and $J_{2,3}/J_0=\beta$, we allow the corresponding couplings on Bob's side to vary according to
\begin{equation}
\frac{J_{N-1,N}}{J_0}=\alpha(1+\delta\alpha), \qquad
\frac{J_{N-2,N-1}}{J_0}=\beta(1+\delta\beta),
\end{equation}
where $\delta\alpha$ and $\delta\beta$ represent relative deviations from the ideal values. This procedure models local imperfections in the couplings at the boundaries of the chain.

\begin{figure*}[htp]
\centering
\includegraphics[width=0.8\linewidth]{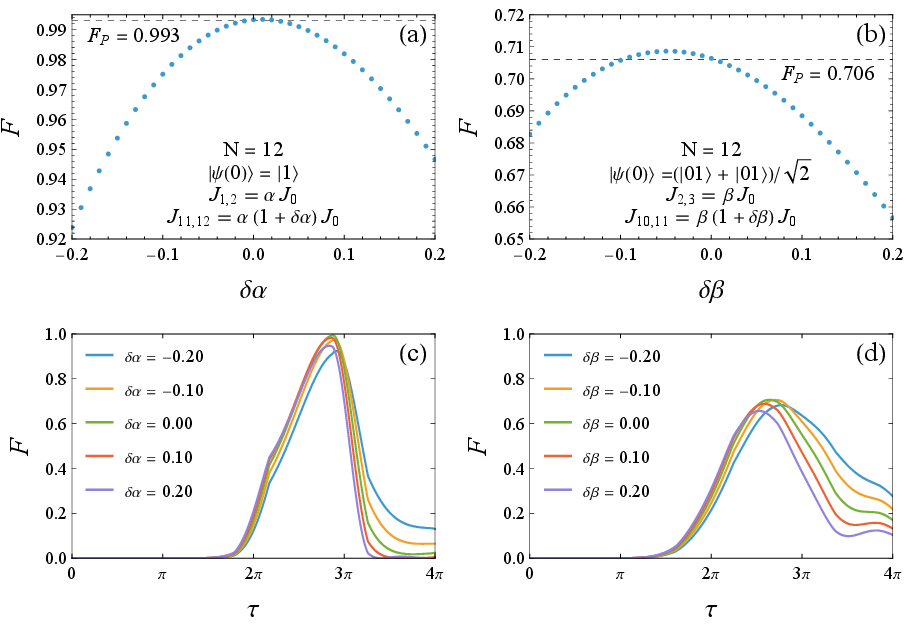}
\caption{Robustness of the transmission fidelity under asymmetric boundary couplings. 
\textbf{a} Variation of the parameter $\delta\alpha$ for the transfer of the state $\ket{\psi(0)}=\ket{1}$ in a chain with $N=12$ qubits. The parameters are $\beta=1$, $\alpha_1=0.47$, $\alpha_2=1.01$, $\omega=1.84$, and $\eta=0.64$. The disorder parameter $\delta\alpha$ varies from $-0.2$ to $0.2$ with increments of $0.01$. The dashed line indicates the ideal symmetric case ($\delta\alpha=0$). \textbf{b} Analogous case for Bell-state transfer $\ket{\psi(0)}=\ket{\psi^+}$, using parameters $\alpha=1$, $\beta_1=0.80$, $\beta_2=1.06$, $\omega=1.78$, and $\eta=0.43$. The disorder parameter $\delta\beta$ follows the same range as $\delta\alpha$. Time evolutions of the transmission fidelities of \textbf{c} $\ket{\psi(0)}=\ket{1}$ and \textbf{d} $\ket{\psi(0)}=\ket{\psi^+}$ for representative values of $(\delta\alpha,\delta\beta)$ as functions of $\tau$}
\label{Fig6}
\end{figure*}

To evaluate the robustness of the protocol, we select representative parameter sets obtained in the optimized Parrondo configurations for both transmission scenarios studied previously. For each case we vary the disorder parameters in the range $-0.2 \leq \delta\alpha,\delta\beta \leq 0.2$, corresponding to up to $20\%$ deviations from the ideal values. The parameters are sampled with increments of $0.01$, resulting in $41$ simulations for each disorder variable. The results are shown in Fig.~\ref{Fig6}. It is important to emphasize that the objective of this analysis is not to search for improved transmission conditions under asymmetric couplings, but rather to evaluate how sensitive the optimized protocol is to experimentally realistic imperfections. We observe that small deviations ($|\delta\alpha|,|\delta\beta|\leq0.02$) do not significantly degrade the optimized transmission fidelity, indicating that the protocol exhibits good structural robustness against small imperfections. However, larger deviations lead to noticeable degradation of the transmission quality, as illustrated in Fig.~\ref{Fig6}a and b. This behavior indicates a moderate sensitivity of the protocol to imbalances in the boundary couplings.

Another relevant observation is that the transmission channel for entangled states is more sensitive to disorder than the single-qubit transmission scenario. This occurs because the Bell-state transmission protocol already starts from a lower baseline fidelity in the ideal case ($\delta\beta=0$), making it more susceptible to parameter fluctuations. Overall, these results show that the Parrondo-based protocol tolerates small experimental imperfections while still maintaining high transmission fidelity.

Lastly, this analysis reinforces the motivation for focusing on the first transmission peak throughout this work. As illustrated in Fig.~\ref{Fig6}c and d, even small deviations in the coupling parameters can significantly modify the subsequent time evolution of the system. Such deviations introduce dispersion and loss of coherence in the quantum channel, making the later transmission peaks increasingly unpredictable in the presence of noise.

\subsection{Temporal noise}

Another important robustness test concerns small fluctuations in the switching times between the two Hamiltonians, i.e., a systematic timing error. To account for this effect, we redefine the time intervals introduced in Eq.~\eqref{eq:periods} as
\begin{equation}
\tilde{T}_1=T_1+\Delta T_1,
\qquad
\tilde{T}_2=T_2+\Delta T_2,
\end{equation}
where $\Delta T_1$ and $\Delta T_2$ represent relative deviations from the ideal switching times $T_1$ and $T_2$. Rewriting $\epsilon_n=\Delta T_n/T_n$, where $\epsilon$ denotes the relative deviation of the driving period from its nominal value, we get 
\begin{equation}
\tilde{T}_1=T_1(1+\epsilon_1),
\qquad
\tilde{T}_2=T_2(1+\epsilon_2),
\end{equation}
where each random variable $\epsilon$ is independently sampled from a uniform distribution over the interval $[-\epsilon_{\max},\epsilon_{\max}]$ at every Hamiltonian switch, with $|\epsilon|\leq\epsilon_{\max}$. Although the random variables associated with the two Hamiltonians are statistically independent and are resampled independently at each switching event, we adopt the convention that both Hamiltonians share the same maximum disorder amplitude, $\epsilon_{\max}$, throughout the simulations.

\begin{figure*}[htp]
\centering
\includegraphics[width=0.4\linewidth]{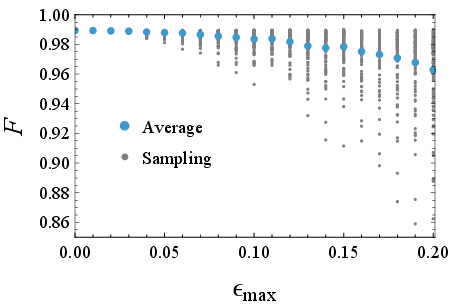}
\caption{Maximum transmission fidelity at the first arrival peak as a function of the relative deviation of the driving period, $\epsilon$, for a chain with $N=10$ qubits. The driving parameters are fixed at $\alpha_1=0.51$, $\alpha_2=1.01$, $\beta=1$, $\gamma=1$, $\omega=1.14$, and $\eta=0.42$. The vertical line indicates the ideal driving period ($\epsilon=0$). The protocol remains robust against small systematic deviations in the driving period, while larger deviations progressively reduce the maximum transmission fidelity}
\label{Fig7}
\end{figure*}

Figure~\ref{Fig7} presents the single-qubit transmission fidelity in the first peak for each value of $\epsilon_{\max}$ in the range $0 < \epsilon_{\max} \leq 0.2$, with increments of $0.01$, 100 independent simulations were performed, resulting in a total of 2000 simulations. The results show that the protocol remains relatively robust against small timing fluctuations, although it is more sensitive to temporal imperfections than to the static variations of the coupling constants discussed previously. This increased sensitivity is expected because the optimal transmission is obtained from careful optimization of the driving frequency $\omega$ and the asymmetry parameter $\eta$. Consequently, even small deviations in the switching times alter the accumulated phases during the Floquet evolution, reducing the constructive interference responsible for the fidelity enhancement. Nevertheless, the protocol maintains high transmission fidelities for timing fluctuations of approximately $1\%-2\%$, indicating that it is sufficiently robust against realistic experimental uncertainties in switching times.
\section{Conclusions}\label{sec:conclusions}

In this work we have investigated the use of Parrondo-like dynamical strategies to enhance quantum-state transmission in spin chains. The protocol consists of periodically alternating between two Hamiltonians that individually produce suboptimal transmission fidelities. Despite each configuration being a losing strategy when applied independently, their time alternation can generate a constructive interference mechanism that improves the overall transmission fidelity.

We have first analyzed the transmission of single-particle states by modulating the boundary coupling parameter $\alpha$ while keeping $\beta=1$. By identifying static configurations that individually yield fidelities lower than the uniform chain ($\alpha=\beta=1$), we have demonstrated that periodic alternation between two such losing configurations can produce fidelities significantly higher than the reference value. This behavior constitutes a clear manifestation of the Parrondo effect in the context of quantum state transfer. The improvement arises in a finite-frequency regime where the effective Hamiltonian generated by the periodic driving differs from the simple high-frequency average predicted by the Floquet-Magnus expansion.

We have then extended the analysis to the transmission of entangled Bell states by varying the coupling parameter $\beta$ while keeping $\alpha=1$. Similarly to the single-particle case, we have identified parameter combinations for which the individual static Hamiltonians yield fidelities below the uniform reference value, yet their periodic alternation produces enhanced transmission. The results show that the Parrondo mechanism remains effective even for the transfer of entangled states, although the entangled transmission channel exhibits a higher sensitivity to parameter variations.

We have also investigated the influence of temporal asymmetry in the driving protocol through the parameter $\eta$, which controls the relative duration of the two Hamiltonians within each driving period. By systematically exploring the parameter space $(\omega,\eta)$, we found that introducing asymmetry can further optimize the transmission fidelity, particularly for longer chains. In several cases the protocol achieved fidelities above $99\%$, even when the static configurations individually perform significantly worse than the uniform chain.

Finally, we have examined the robustness of the protocol against local disorder in the boundary couplings. By introducing relative deviations of up to $20\%$ in the coupling strengths on Bob's side of the chain, as well as systematic errors in the switching period, we show that the protocol tolerates small imperfections without substantial degradation of the optimized fidelity. Larger deviations, however, lead to noticeable losses in transmission quality, indicating a moderate sensitivity of the protocol to boundary asymmetries. Importantly, the first transmission peak remains relatively stable, reinforcing its use as the primary indicator of transmission performance. 

In addition to static asymmetries in the boundary couplings, we have examined the effect of small temporal deviations in the driving period. These fluctuations model timing jitter in the switching protocol. The results show that the fidelity remains stable for sufficiently small deviations, indicating that the Parrondo enhancement does not rely on infinitely precise timing. A full treatment of random bulk disorder would require averaging over disorder realizations and would introduce additional localization effects, such as Anderson localization. A detailed analysis of these effects is beyond the scope of the present work and is left for future study.

Overall, our results demonstrate that Parrondo-like dynamical strategies provide an effective mechanism for enhancing quantum-state transmission in spin chains. Beyond their conceptual interest, these protocols offer a promising route for improving state-transfer fidelity in short quantum communication channels where static Hamiltonian engineering alone may be insufficient. Future work may explore extensions of this approach to longer chains, higher-excitation sectors, different spin models, and experimentally relevant platforms where time-dependent control of coupling parameters is available. It would also be interesting to investigate the large-system scaling of the Parrondo enhancement, as well as the existence of upper bounds for the fidelity enhancement and optimal driving protocols, in order to better understand the fundamental limits of this approach.

\section*{Acknowledgements}
R. Vieira thanks Fundação de Amparo à Pesquisa e Inovação do Estado de Santa Catarina (FAPESC) for funding. E. P. M. Amorim thanks J. Longo for her careful reading of the manuscript.

\section*{Author contributions}
E.P.M.A. and R.V. contributed equally to this work. Both authors jointly conceived the research problem, developed the theoretical framework, performed the numerical simulations, analyzed and interpreted the results, and wrote and reviewed the manuscript. All authors approved the final version of the manuscript.

\section*{Funding} 
This work was supported by Conselho Nacional de Desenvolvimento Científico e Tecnológico (CNPq) through grant number 409673/2022-6. R. Vieira has received research support from Fundação de Amparo à Pesquisa e Inovação do Estado de Santa Catarina (FAPESC), Edital 25/2025.

\section*{Data availability}
All data generated or analysed during this study are included in this manuscript.

\section*{Declarations}

\subsection*{Ethical approval and consent to participate}
Not applicable. This study did not involve human participants, human data, or animals.

\subsection*{Consent for publication}
Not applicable.

\subsection*{Competing interests}
The authors declare no competing interests.

\end{document}